\documentclass[12pt]{article}
\usepackage{graphics}
\input epsf
\usepackage{amsmath,color,fancybox,graphics,graphpap,rotating}
\definecolor{yellow}{rgb}{0.95,0.75,0.1}
\definecolor{red}{rgb}{1,0,0}
\definecolor{darkred}{rgb}{0.5,0,0}
\definecolor{green}{rgb}{0,1,0}
\definecolor{blue}{rgb}{0,0.5,1}
\definecolor{bgcolor}{rgb}{0.9,0.9,0.999}

\newcommand{\be}{\begin{eqnarray}}
\newcommand{\ben}{\begin{eqnarray}\nonumber}
\newcommand{\ee}{\end{eqnarray}}
\newcommand{\nee}{\nonumber \end{eqnarray}}

\begin{document}
\title{
\begin{flushright} 
UAHEP062
\end{flushright}
\vskip 1cm
Neighboring Valley in the String Landscape\\
A Phase Transition to Exact Susy
\footnote{ 
talk presented at the Susy06 conference, 
Irvine, CA, June 12-17, 2006}}
\author{L. Clavelli\footnote{lclavell@bama.ua.edu}\\
Department of Physics and Astronomy\\
University of Alabama\\
Tuscaloosa AL 35487\\ }
\date{July 3, 2006}
\maketitle
\begin{abstract}
The observation in the universe of a small but positive vacuum energy 
strongly suggests, in the string landscape picture, that there will 
ultimately be a phase transition to an exactly supersymmetric universe.
This ground state or "true vacuum" of the universe could be similar to
the minimal supersymmetric standard model with all the susy breaking
parameters set to zero.  Alternatively, it might be similar to the
prominent superstring theories with nine flat space dimensions or to
the supersymmetric anti-deSitter model that seems to be equivalent to
a conformal field theory.  We propose that the
dominant phenomenological feature of these potential future universes
is the weakening of the Pauli principle due to Fermi-Bose degeneracy. 
Providing the phase transition occurs in the cosmologically near future,
an exact supersymmetry could extend the life expectancy of intelligent
civilizations far beyond what would be possible in the broken susy
universe.
\end{abstract}
\renewcommand{\theequation}{\thesection.\arabic{equation}}
\renewcommand{\thesection}{\arabic{section}}
\section{\bf Introduction}
\setcounter{equation}{0}
  
   Several recent observations have made it increasingly likely that 
the expansion of the universe is accelerating in a way consistent with an
interpretation in terms of a positive vacuum energy density of approximate
magnitude
\be
         \epsilon_{now} = 3560 MeV/m^3 = (.0023 eV)^4 \quad .
\label{vacenergy}
\ee
This is some 124 orders of magnitude greater than 
the natural value that might have been expected for this quantity:
\be
      M_{Planck}^4 = 10^{127} MeV/m^3 \quad .
\ee

\begin{figure}[ht]
\begin{center}
\epsfxsize= 4.5in 
\leavevmode
\epsfbox{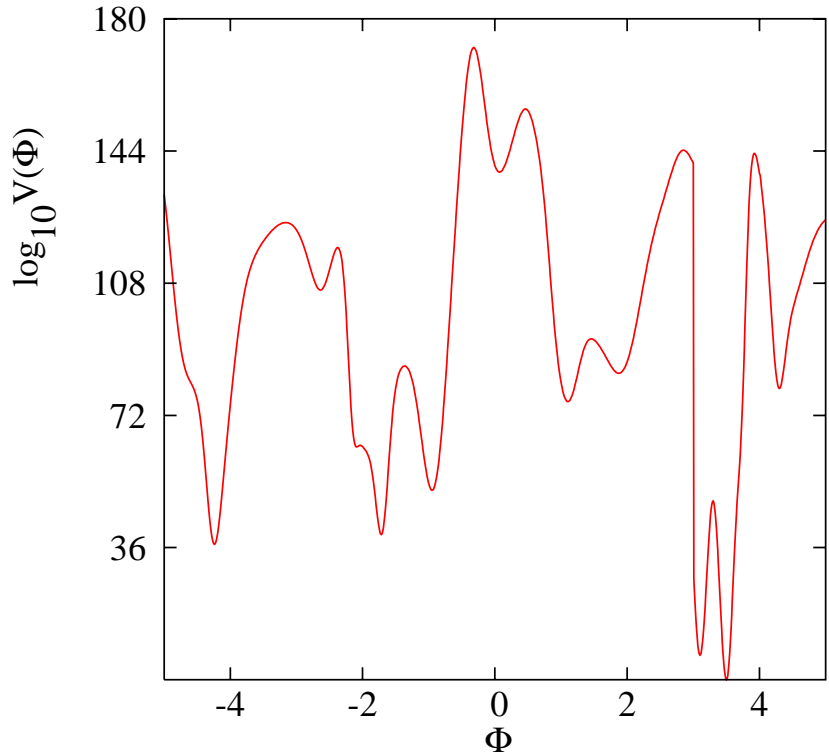}
\end{center}
\caption{A schematic representation of the effective potential in the string landscape
picture.  The potential measured in units of MeV/m$^3$ is not drawn to scale. 
The y axis has a broken scale taken to be linear in V at low values of the potential.
Our world with a small vacuum energy is shown together with the neighboring
exact susy phase with zero vacuum energy.}
\label{landscape}
\end{figure}

We assume that, in addition to our broken susy universe,  there is a neighboring
valley in the string landscape described by a perfect supersymmetry (susy)
\cite{future} and, most likely, a vanishing cosmological constant as pictured in
figure \ref{landscape}.  For our current purposes it matters little whether
the susy minimum has four or more dimensions or whether the space is
flat, deSitter, or anti-deSitter as long as the cosmological constant is
not much greater in absolute value than our current one.  
We postulate that this susy minimum is the true vacuum
and, therefore, the final phase of the universe.  At CERN, the broken susy
phase has been referred to as "Susonia" and we have correspondingly
suggested the future exact susy phase be called "Susyria".  

In this talk we discuss the
properties of this end-phase and address the four basic questions that were 
raised in \cite{future}.  While our primary interest, at present, is in the final transition from our broken susy world to the exact susy universe, it is thought 
that
the inflationary phase in the very early universe corresponded to a sequence of
similar phase transitions to progressively lower vacuum energies.  Many such
scenarios have been considered recently by Susskind and collaborators 
\cite{Susskind} as well as by others.  We do not consider here the 
disfavored possibility of quantum jumps back to higher local minima.
Some, presumably for philosophical reasons,  have pursued the idea of an "eternal 
inflation" continually throwing off bubble universes in some of which the cosmological
constant is small enough to support the evolution of life.  Others, for similar
reasons, have embraced the idea of a
cyclical universe with repeated big crunches alternating with big bangs.   
These alternative philosophies envision an infinite number of life-supporting
universes existing elsewhere in space time causally disconnected from our
world.  They therefore seem uneconomical in the extreme although this might
depend on one's philosophical presuppositions. 
As, possibly, the most economical interpretation of the big bang data we prefer
an absolute beginning of the multiverse at a finite time in the past in a
state near the peak of the vacuum energy distribution although this is not
crucial to the current discussion. We do assume that immediately after the
big bang the universe was in a local minimum of vacuum energy density near $M_P^4$
and was, therefore, inflating rapidly.
We assume that
the distribution of string minima in vacuum energy density is strongly peaked at
this natural value $M_{P}^4$ as, for example, in a Gaussian distribution:
\be
     N(\epsilon) = N_0 exp^{-k (\epsilon - M_P^4)^2/M_P^8} \quad .
\label{Nofepsilon}
\ee 

If $k$ is large enough, the vast majority of local minima are of order $M_P^4$
but if the total number of minima proportional to $N_0$ is also very large,
a few of the minima will have vacuum energy density values
$\epsilon$ below the maximum at which life could evolve.  
This maximum is about two orders of magnitude higher than our observed vacuum
energy density \cite{Weinberg}.
The existence of
such a mildly accelerating universe is the first prerequisite for the study
of physics by intelligent beings.   String theory suggests
that the distribution of eq. \ref{Nofepsilon} integrates up to a total number of
local minima above $10^{100}$.  However, it is not
enough to have a local minimum with a small enough vacuum energy.  It is also
crucial for the rise of life that the transition to our minimum $\epsilon_{now}$ 
happen in a 
time that is neither too short nor too long.  If the transition takes too long
the universe would be too dilute to form galaxies, planets, and life.  If the
transition takes place too rapidly, there would be too sudden an entropy
growth and too much overheating of the nascent universe.  Susskind and collaborators
\cite{Susskind} have investigated many scenarios for the emergence from the 
inflationary era into the present mildly accelerating universe.   

\begin{figure}[ht]
\begin{center}
\epsfxsize= 4.5in 
\leavevmode
\epsfbox{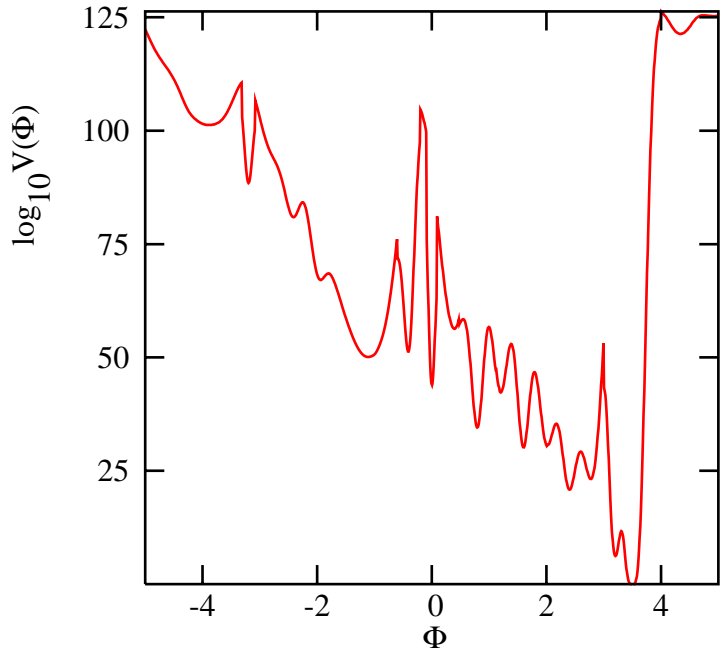}
\end{center}
\caption{A schematic representation of the effective potential in the string landscape
picture with re-ordered minima to reflect the slow roll of the effective field down
to the current broken susy minima and showing the neighboring exact susy true vacuum.  
}
\label{landscp6}
\end{figure}

From Coleman and collaborators \cite{Coleman}, we adopt the simplest vacuum decay
probability per unit time per unit volume from a minimum of initial energy density
$\epsilon_0$ to a lower minimum of energy density $\epsilon$.  Multiplying this by
eq.\,\ref{Nofepsilon}, the number of minima near $\epsilon$, gives us the transition rate per unit volume
\be
    \frac{d^2P}{dt dV} = N(\epsilon) A e^{-13.5 \pi^2 S^4/(\epsilon_0 - \epsilon)^3} \quad .
\label{decayprob}
\ee
Here $A$ is an undetermined normalization and $S$, the surface tension of the bubble of vacuum $\epsilon$, is a function of the initial and subsequent vacuum energies.  
If $S$ is sufficiently small and/or $k$ is sufficiently large, the peak of the
transition probability as a function of $\epsilon$ is to values of $\epsilon$
very near $\epsilon_0$.  This leads to a ``slow roll'' of the effective scalar field
$\Phi$ down to our universe.  If we re-order the minima in fig.\,\ref{landscape}
the picture suggested is as appears in fig.\,\ref{landscp6} where our universe with a
broken susy and a residual vacuum energy is shown together with the ultimate 
exact susy true vacuum.  If the true vacuum is a deSitter space with negative
vacuum energy density many of our considerations will remain true although the
universe will ultimately collapse in a big crunch.  
If string theory is a guide \cite{Giddings}, a future transition to some such
universe is essentially inevitable.
Since the particle masses
, as well as other properties of the theory, are different in each of the meta-stable intermediate universes, the entropy 
release is not necessarily as severe a problem as in the conventional theory.
In the exact susy true vacuum susy particles will be degenerate with their standard
model (SM) counterparts.  We will assume the common masses are those of the SM
particles in our broken susy world.  For simplicity we can think of the minimal
supersymmetric standard model (MSSM) with all of the susy breaking parameters put
to zero.  Also, at the current stage of investigation, we assume the future 
topology of space time is as in the broken susy world.

\begin{figure}[ht]
\begin{center}
\epsfxsize= 4.5in 
\leavevmode
\epsfbox{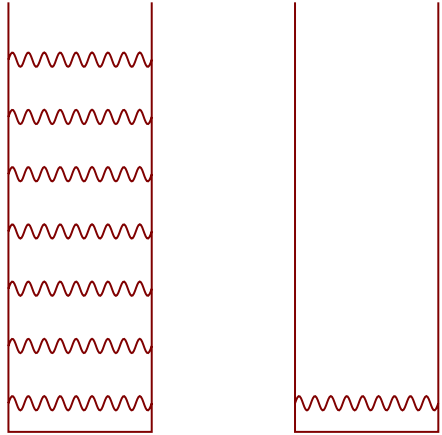}
\end{center}
\caption{A Fermi degenerate system in the broken susy phase (on left) and after
a phase transition to exact susy (on right).
}
\label{Pauli}
\end{figure}

     We have proposed \cite{future} that the primary distinguishing property
of matter in the exact susy phase relative to our universe 
is an effective weakening of the Pauli Principle.   This is due to the fact that,
in the broken susy world, every atom above Helium is characterized by energy
permanently stored in a Pauli tower of electrons and in a separate tower of
nucleons in the atomic nucleus.  In exact susy, conversion of Fermion pairs to
degenerate scalar pairs not governed by the Pauli principle allows the release
of this energy.
\be
     f f \rightarrow {\tilde f}{\tilde f}
\label{pairconversion}
\ee
This process \cite{CP} occurs in every susy model with or without $R$ parity violation.
For electrons the pair conversion process is mediated by photino exchange while
for quarks it is mediated also by gluino exchange.  Thus, following a phase transition
to exact susy, fermions in excited states will convert in pairs to bosons which can
then drop into the ground state as indicated in fig.\,\ref{Pauli}.

Susy atoms in their ground state will therefore
consist of zero, one, or two fermionic electrons with the rest of the leptonic cloud
consisting of selectrons in the ground state.  Similarly, all the particles in the
nucleus will occupy the ground state wave function with as many as necessary being
scalars.  There will be no orbital angular momentum in ground state susy nuclei
or leptonic clouds and therefore greatly restricted magnetic moments.

\begin{figure}[ht]
\begin{center}
\epsfxsize= 4.5in 
\leavevmode
\epsfbox{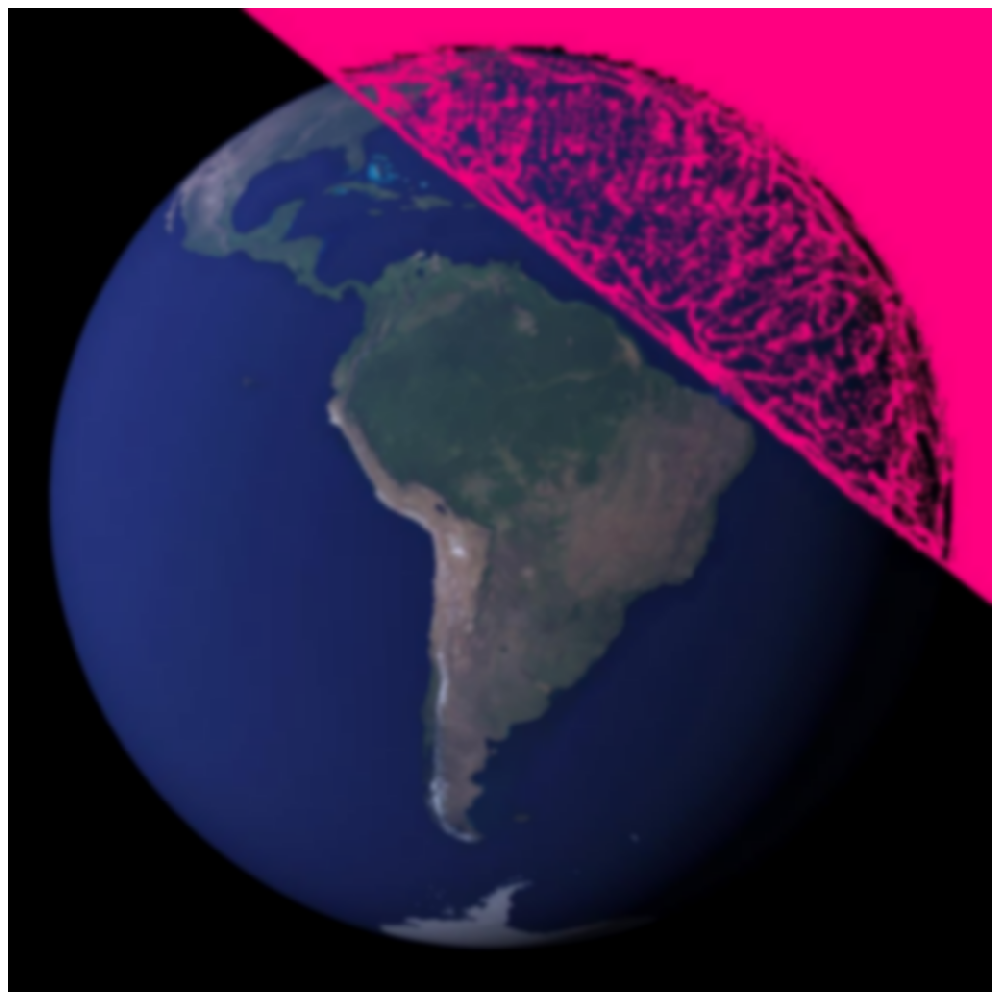}
\end{center}
\caption{}
\label{earth}
\end{figure}

A phase transition in vacuum will begin with the nucleation of a bubble of 
true vacuum with radius greater than a critical radius
\be
       R_c = \frac{3S}{\epsilon_{now}}  \quad .
\ee
Although he did not specifically consider a supersymmetric true vacuum, 
the work of \cite{Coleman} generically predicts that such a bubble will expand 
in the vacuum at the 
speed of light converting all the matter in its path to the new phase.  As such, the
bubble wall will strike each planet without the possibility of advance warning.
An artist's depiction of such a bubble striking the earth might be as in 
fig.\,\ref{earth}.  Although there can be no advance warning of the arrival time 
of a susy bubble nucleated in the vacuum,
the inevitability of such a phase change is implied if gamma ray bursts
or other violent astrophysical events are due \cite{CK} to density stimulated
susy phase transitions in degenerate stars (for a review see ref.\,\cite{origin}).
Stimulated phase transitions are confined to the region of high density although
photons, light in both phases, can escape.  After such transitions, the absence of
an outward degeneracy pressure will lead to gravitational collapse to a susy 
black hole.  

{\bf The four basic questions posed in ref.\,\cite{future} are}
\begin{enumerate}
\item{\bf Could life have arisen if there had been a phase transition directly from the
inflationary era to the exact susy minimum?}
 
Only if other
low lying minima in the string landscape are few in number or are unsuitable for the
evolution of life does observer bias or the anthropic principle
provide some understanding of why the universe is as it is.  
The likelihood that exact susy minima exist in string theory forces
one to examine the possibility of life arising after a transition from the inflationary
era direct to an exact susy minimum.  There are several weak hints arguing that no such possibility exists.  First of all, one could note that galactic evolution seems to rely on a large dark matter component to provide the gravitational well within which normal matter
can condense into galaxies.  In the broken susy world, this role is played by the stable
lightest supersymmetric particle (LSP) which is thought to have a mass in the $100$ GeV
region.  In the exact susy phase, the LSP's are massless partners of the photon and
graviton and there is no heavy stable particle to provide dark matter.  Of course one
could investigate whether other non-susy particles such as heavy neutrinos or axions
could plausibly provide the dark matter in the exact susy phase without otherwise
hindering the evolution of life.  Other possible problems with production and
distribution of heavy elements following a transition directly from the inflationary
era to the exact susy phase have been noted in ref.\,\cite{future}.

Thus it ‎is possible that the existence of life in a susy universe might require the
prior generation of heavy elements in a broken susy phase.

\item{\bf Could life survive, or re-establish itself, following a transition from our
broken susy world to the exact susy world?}
 
It is easy to find possible impediments, such as the one discussed in point $1$ 
above, to the evolution of life in an exact susy universe following a direct transition from an inflationary era.  If it is confirmed that these are incurable, one could still
ask whether an exact susy universe could support life if there was an intermediate
broken susy phase.  Like the time critical property of the transition from the
inflationary era to our calm broken susy universe, the transition to exact susy 
might also be time critical.  If the current accelerating phase lasts too long,  most stars will consist of white dwarfs out of causal contact with each other.
At that point
it is unlikely that life could be revived through a susy phase transition.
The energy release in the conversion of fermions to bosons would be primarily
in the form of gamma radiation and would not efficiently redistribute heavy 
nuclei through the universe.  On the other hand, if the transition takes place
while there are still earth-like planets orbiting burning stars, it is 
conceivable that life could re-establish itself.  Although the radiation released
from the Pauli towers would totally sterilise planets, it is not sufficiently
energetic to totally dissociate nuclei.  Leptons would eventually condense
on heavy nuclei and it is plausible that molecular binding
qualitatively similar to that of our world would occur.
Afterwards, as we discuss in point 3
below, heavy elements would beta decay down to susy nuclei near Oxygen.  
Since all the
elements needed to form DNA and $96\%$ by weight of animal species are no heavier 
than Oxygen, 
evolution would be expected to recur leading to the re-emergence of species 
qualitatively similar to many of those in the broken susy world and defined by
the same genetic codes.

The properties of bulk susy matter are discussed in point 3 below and in point 4
we show that the expected time of the transition is close to the critical time
discussed here for the re-establishment of life.

\item{\bf What would be the primary characteristics of the physics (and biology, if any)
of the exactly supersymmetric phase?}
 
The primary distinguishing features of bulk susy matter relative to matter in the
broken susy phase are the greater numbers of states due to supersymmetry and the
weakening of the Pauli Principle due to the possibility of pair conversion from
Fermions to Bosons according to eq.\,\ref{pairconversion}.  Whenever, in the
broken susy phase, bound Fermions are forced into elevated energy levels, in the
susy phase it will be advantageous for them to convert in pairs into their
degenerate susy partners which, being Bosons, can drop into the ground state.
Susy atoms will consist of zero, one, or two fermionic electrons.  The remaining
leptonic cloud will consist of selectrons.  The entire leptonic cloud will be in
the 1S state.  This has the effect of making susy atoms much smaller in general
than their broken susy counterparts although the effect is moderated by the
increased self repulsion of the cloud.  Smaller atoms in a solution will be 
expected to have slower reaction rates due to the decreased probability of 
collisions but might bind more tightly into molecules because of the smaller
intra-molecular distances.
    
    Susy nuclei would be expected to be sneutron rich since the increased binding
with extra sneutrons would not be in competition with the Pauli exclusion principle
which forces Fermionic neutrons into higher energy levels.  Sprotons are also
unaffected by the Pauli principle but their number is limited by Coulomb repulsion.
In ref.\,\cite{future} we have considered the beta decay constraints on snuclear
stability:
\be
   \left( \frac{2 a_c (Z-1/2)}{M_n - M_p + m_e}\right)^2 < A < 
\left( \frac{2 a_c (Z+1/2)}{M_n - M_p - m_e}\right)^2
\ee
where $a_c$ is the coefficient of the Coulomb term in the semi-empirical
mass formula for nuclei.  We have assumed that the interaction strengths
are similar to those in the broken susy world.  Assuming degenerate
susy multiplets have the same masses as the standard model particles in
the broken susy world, the atomic weight of snuclei increases rapidly with
atomic number so that stable elements above susy Oxygen must have atomic weights
well above $238$.  Since in the broken susy world there are long-lived elements
with atomic weights only up to this number, after a susy phase transition
only elements up to susy Oxygen would be abundant.  The elements with higher
atomic number would beta decay down to Oxygen and below.
A brief period of fusion
burning might rebalance relative abundances of the light elements without leading to 
appreciable quantities of elements beyond Oxygen due to the requirement that
higher elements have prohibitively large numbers of sneutrons.
   
     As previewed in point $2$, the constituents of life would then be available 
and, assuming molecular binding is cooperating, life forms similar to many of
those we are familiar with would inevitably evolve given enough time.  This 
assumes that the trace elements heavier than Oxygen found in living systems can
be somehow dispensed with or replaced with lighter elements.

     Fusion in susy phase stars will proceed beyond the iron limit
of the broken susy phase because of the absence of a restrictive Pauli Principle.
This means that susy stars will burn considerably longer than normal stars.
Unless some new considerations come into play, they will, of course, 
eventually exhaust their fuel and collapse into black 
holes irrespective of their mass since there will be no degeneracy pressure
to prevent collapse. 

\item{\bf Can we estimate the probable time remaining before our universe converts to
a susy world?}
 
The vacuum decay probability per unit time per unit volume as given in eq.\,\ref{decayprob} depends on the vacuum energy of the current phase, eq.\,\ref{vacenergy}. 
Thus the transition rate is proportional to the volume in which a phase change is
possible.  This volume is proportional to the cube of the scale factor in the
Friedman-Robertson-Walker metric which, for positive cosmological constant, is
exponentially growing at large times. 

In the presence of a vacuum energy density, $\epsilon$, the scale 
factor of general relativity will satisfy
\be
     \frac{\ddot{a}}{a} = - \frac{4 \pi G_N}{3} (\rho_{vac} + 3 p_{vac}) \quad .
\ee
Putting $p_{vac}=-\rho_{vac}= - \epsilon$, where $\epsilon$ is our current vacuum
energy, the $\epsilon_{now}$ of eq.\,\ref{vacenergy}, this has the solution
\be
     a(t) = e^{\gamma t/3}a(0)\left (1 + (\frac{3 \dot{a(0)}}{\gamma a(0)}-1)\frac{1-e^{-2 \gamma t/3}}{2} \right )
\label{scalefactor}
\ee
where, in terms of Newton's constant, $G_N$,
\be
     \gamma = \sqrt{24 \pi G_N \epsilon} \quad .
\ee
Neglecting sub-leading terms, we may write 
the volume of the universe at time $t$ in terms of its present volume $V(0)$ as
\be
       V(t) = V(0) e^{\gamma t} \quad .
\ee
The natural time scale for the growth in volume of the universe is
\be
    \gamma^{-1} = 5.61 \cdot 10^9 {\displaystyle {yr}} \quad .
\label{timescale}
\ee
This time is comparable to the current age of the sun and to
its expected additional lifetime before becoming a red giant. 
The volume of the universe is at least as big as the Hubble volume
\be
       V(0) > V_H = 7.79 \cdot 10^{78} m^3 \quad .
\ee
What is the probability that such a bubble will strike Earth or
some other location in a given time from now?
Once nucleated somewhere in the universe, the bubble will
require some time to propagate to any particular location such as 
that of Earth.

The probability per unit time for a susy bubble to arrive at any given location
at local time t
is the probability per unit time for
a critically sized bubble to be nucleated at any position $r'$ at the
retarded time $t' = t - r'/c$
\be
    \frac{dP(0,t)}{dt} = \int d^3r' e^{\gamma t'} \frac{dP(r',t')}{dV' dt'}
          dt' \delta(t' - t + r') = e^{\gamma t} A e^{-B} \int d^3r' e^{-\gamma r'}
  \quad .
\ee
This can be written
\be
    \frac{dP(0,t)}{\gamma dt} = e^{(\gamma t - B + \ln{(8 \pi A/\gamma^4)})} \quad .
\label{transprob}
\ee

An integrated probability over any time
interval exceeding unity should be interpreted as the probable number of
susy bubbles hitting the earth in that time interval.  
We know that this has not happened between the time of the big bang and now.
Requiring that the integrated probability from the big bang to now ($t=0$)
be less than unity suggests 
\be
         B > ln(8 \pi A/\gamma^4) \quad .
\label{Blimit}
\ee
If we allow ourselves to
consider saturating the limit \ref{Blimit}, 
there is a non-negligible probability that the Earth will be swallowed 
by a susy bubble in a time $T$ from today that is smaller than $1/\gamma$.
This can be seen by replacing the inequality of eq.\,\ref{Blimit} by an
equality and integrating eq.\ \ref{transprob} from $0$ to $T$:
\be
     P(T) = e^{\gamma T} - 1 \quad .
\ee
This is only relevant while $P(T)<1$ since the collision of multiple susy 
bubbles with Earth is overkill. 

We find it somewhat amazing that the natural time scale defined by the 
observed vacuum energy eq.\,\ref{timescale} is at the boundary between
that at which the re-evolution of life is possible and that at which
a susy phase transition would lead to a lifeless universe of isolated
susy black holes as described in point $2$ above. 
\end{enumerate}

\section{\bf Conclusions}
\setcounter{equation}{0}
\par

     The time $\gamma^{-1}$
provides an approximate upper limit to the lifetime of intelligent
species in our broken susy universe.  At times significantly larger
than this all earth-like planets will have been devoured by red giants
or obliterated in supernovae.  $99\%$ of stars will be in the form of
cold white dwarfs accelerating rapidly away from each other.  
The others will be in the form of neutron stars or black holes.
This corresponds 
to the much-discussed heat death of the universe, an end in ice.  
Alternate cosmologies
under consideration postulate a reversal of the current outward
acceleration of the universe toward an ultimate "big crunch", an
end in fire, or toward
a recurring eternal sequence of big bangs followed by big crunches.   

We have outlined a possible new end phase scenario, a phase 
transition to an exactly supersymmetric universe.
Because of the outward acceleration of the universe, the transition
probability per unit time is an exponentially increasing function
of time.  Providing the inevitable transition occurs before 
about $\gamma^{-1}$ there is a possibility that supersymmetric life forms
could evolve.  

  This work was partially supported by the DOE under grant number
DE-FG02-96ER-40967.  We acknowledge stimulating discussions with
Paul Cox, Irina Perevalova, Tim Lovorn, and Stephen Barr.

\end{document}